\documentclass[aps,prd,twocolumn,showpacs,superscriptaddress,groupedaddress]{revtex4}
\usepackage{amsmath}
\usepackage{amssymb}
\usepackage{graphicx}%

\begin{document}

\title{New $B^0_s \pi^{\pm}$ and $D^{\pm}_s \pi^{\pm}$ states  in high energy multi-production process}

\date{\today}
\author{Yi Jin$^1$,   Shi-Yuan Li$^2$  and Shu-Qing Li$^2$\\
{\small $^1$ School of Physics and Technology, University of Jinan, Jinan 250022,
	P. R. China}\\
{\small $^2$ School of Physics, Shandong University, Jinan 250100, P. R. China}
}

\begin{abstract}
The  production rate of the X(5568) measured by D0 collaboration is quite large and  difficult to be understood by various general hadronization mechanism. We propose the inclusive  production formulation for the cross section, and predict the distributions and production rates of X(5568) at LHC energies, which are crucial information for the relevant
measurements. We also suspect $D^{\pm}_s \pi^{\pm}$ state can be copiously produced and observed. 
	

\end{abstract}

\maketitle

Recent years one witnesses the observation of exotic multi-quark hadrons \cite{bela,belb,cdfa,belc,lhcba}. 
  The flavour number presented in the exotic  hadron exceeding three is very crucial in clearly identifying the valence
quark number. 
 For a hadron with non zero quantum  numbers as  bottom or charm besides the  isospin ($I_3$) and strangeness,
 one has no other choice than to introduce more than three quarks carrying them.
Exotic multi-quark hadron with only `light' (u, d, s) quarks  might have produced
 in various processes,
but it is not  easy to confirm the valence quark number. {\it Light} Pentaquark is special, but yet of no confirmed evidence
of its production.

 Another significant fact should also be paid great attention on    is that
the  confirmed multi-quark states were all  produced  from  heavier hadron decays.
%
%
 However, this situation may be changed:
The D0 collaboration just
  found a new $B^0_s \pi^{\pm}$ state, X(5568) \cite{D0:2016mwd}, with {\it four flavours}
in this hadron, ---if it is really a particle. This  will  be   quite remarkable, as the first solid evidence of multi-quark
state {\it directly  produced in the   multi-production process of high energy collision}, rather than   from   hadron decays.
  The bottom flavour here is very decisive. Because of the bottom flavour and the  mass,  
it is hardly possible produced from decay of a heavier hadron
(For the charm sector, however, one has  to distinguish prompt ones from those from bottom decay).


 %
 %

  This observation of the X(5568)  \cite{D0:2016mwd} surely turns on
  the significant new page in the study of the multi-quark states.  
   Now one can study their  {\it production mechanism}
   which is tightly related with the inner structure of the relevant hadrons \cite{long1, former},
 {\it in  multi-production processes} at high energy colliders like  Tevatron and LHC.
 It is noticeable that
  in the high energy multi-production processes, the space-time evolution of the quark system and
   the produced hadron system are quite different from that of the decay processes. Besides the different partonic processes and
  kinematic phase space distributions,
  the final state interaction (if existing) is also  different. So comparing the production of the `exotic hadron' in the
  decay and  multi-production processes
   can eliminate the misleading information  possibly rendered by, e.g., phase space effects and  final state
  interactions, etc.

  As a matter of fact, the  measurement  \cite{D0:2016mwd} has  provided the    important  information of the production.
   The  ratio $\rho$ of the yield of   $X(5568)$ to
the yield of the  $B_s^0$ meson in  two kinematic ranges,
10 $< p_T (B_s^0) <$ 15 GeV/$c$ and 15 $< p_T (B_s^0) <$ 30 GeV/$c$, is measured.
The results for $\rho$ are  (9.1 $\pm$ 2.6 $\pm$ 1.6)$\%$ and
(8.2 $\pm$ 2.7 $\pm$ 1.6)$\%$, respectively, with an average of (8.6 $\pm$ 1.9 $\pm$ 1.4)$\%$.
  Here we assume $B^0_s \pi^{\pm}$ is the dominant decay mode of this new state.
This large production rate itself first of all excludes the possibility of decay from heavier particles
like 
$B_c$, 
 which are difficult to produce.
 In this paper we investigate  the production mechanism of this new $B^0_s \pi^{\pm}$ state, taking it as a particle, say,  X(5568).  We emphasize that its  production rate is quite large, 
  and difficult to  be understood by various general hadronization models.
  So we propose the inclusive resonance production formulation
to calculate its  cross sections
 for various collision processes as well as
 energies,
which  provides useful information for the relevant detections, e.g., those at LHC.
We  demonstrate that the transverse and longitudinal momentum distributions, together
with the special property of the decay channel are important for the set of the triggers and data
selection requirements.
 When  just the bottom quark is replaced by the charm quark in X(5568),
assuming   the  structure  hence the production
mechanism for the corresponding four-quark charm  hadron do not change,  the production of the `new $D^{\pm}_s \pi^{\pm}$ state'
is studied.   

As  mentioned above, the production mechanism is
tightly related with the inner structure of the relevant hadron \cite{long1, former}. Our
analysis starts  from assuming X(5568) as four quark state and directly produced together with other general
mesons and baryons, employing  various  hadronization mechanism \cite{Jin:2010wg}.





1) \texttt{String fragmentation model \cite{Andersson:1998tv}}.
The bottom quark  can only be produced by hard scattering between partons. In the hadronization process it links with  another antiquark $\bar q $ by a string. 
 Then this string fragments into several hadrons, among which are the bottom hadrons, e.g.,
 $B^0_s$.
   It is well known that the relative probabilities of the creation of
light flavours $u,d,s$ from vacuum are $1:1:\lambda$, with $\lambda \sim 0.3$ \cite{Agashe:2014kda, Jin:2010wg}.
These values can be extracted  by
looking into the production ratios, e.g.,  $B_s/B_0 \sim 0.3$ in various high energy multi-production processes.
For the investigation here, we also need the production of diquark.
Since we only want to estimate the production rate, not concentrating on correlations,  we adopt the diquark mechanism rather than the popcorn mechanism for simplicity. And this will be consistent with the X(5568) structure when considered as diquark antidiquark pair.
From the production ratios such as $\Lambda_b/B_s \sim 1/2$, we take it as $\lambda/2$. However, if the diquark has a strange flavour, e.g., $us$, a further factor $\lambda$ is introduced.
Here is the relative probabilities:
\begin{eqnarray}
  ~& u:d:s:ud:uu:dd:us:ds:ss \\ \nonumber
= & 1:1:\lambda:\frac{\lambda}{2}:\frac{\lambda}{2}:\frac{\lambda}{2}:\frac{\lambda^2}{2}:
\frac{\lambda^2}{2}:\frac{\lambda^3}{2}.
\label{rostring}
\end{eqnarray}
In string model, quarks and diquarks are created from vacuum  in pair, so here `$u$' means `$u \bar u$', `$ud$' means `$ud \overline{ud}$',
etc.
These relative ratios are consistent with the production rates of $\Xi $, $\Sigma $,  non-strange baryons, as well as their
 relative production ratios to the relevant mesons in a jet
\cite{Agashe:2014kda}.

On the other hand, the production of X(5568) needs a special string: The $bu$ or $bd$ diquark as one of the end of the string
connecting another quark as the other end \cite{Jin:2013bra, Jin:2014nva, Jin:2015mla}. Needless to say, this kind of colour connection is suppressed. We set an ansatz
that this special kind of string $bq -q$, relative to the usual $b-\bar q$ one, receives a suppression factor $\lambda/2$.
This is reasonable since when the strings are `short' enough and they just correspond to a single particle, the suppression
factor just gives the correct relative production rate.
 Hence we estimate the production ratio $X(5568)/B^0_s$ in the string model as:
 \begin{equation}
  P_X= \frac{2 \times \lambda/2}{D_1}
  \times  \frac{\lambda^2/2}{D_2}.
 \end{equation}
 Here $D_1=1+1+\lambda+\lambda/2+\lambda/2+\lambda^2/2$, and $D_2=1+1+\lambda+\lambda/2+\lambda/2+\lambda/2+\lambda^2/2+\lambda^2/2+\lambda^3/2$.
  \begin{equation}
  P_{Bs}= \frac{1+1+\lambda}{D_1}
  \times
  \frac{\lambda}{D_2} ;
 \end{equation}
 \begin{equation}
  \rho_{string}= \frac{P_X}{P_{Bs}} \sim 2 \%.
 \end{equation}
 This is much smaller than the average $8.6 \%$. Even,  we emphasize that the decreasing nature
  with respect to the
 increasing of the transverse momentum indicated by
 the experiment (we will clarify that it must be)
 of the above relative ratio  $\rho$  renders that the total production ratio is larger than $8.6 \%$.
 Furthermore, we ignore  the fragmentation into several kinds of hadrons
 with similar flavour but different spins, which  will even decrease the ratio.
 To get a higher production rate, one needs to tune the relative ratios in Eq. (\ref{rostring}), i.e., to
 increase the relative ratio of the diquarks containing  strange quark(s). However,  similar
 problems as in other models \cite{lsy} to give correct production rates of the  strange baryons will be raised.

2) \texttt{cluster  model and final state interaction \cite{Webber-cluster}}.
In cluster model, we have to introduce a   free parameter to describe the probability  the cluster with mass large
enough  to decay
into X(5568). The value of the parameter can be obtained by fitting the data.  So it is not very meaningful to say it is large or small.
By the help of the recent work \cite{fi},
we find that if  this peak is not a real particle but some kind of final state interaction effect, the observed
`rate' is also very small. The reason is that this kind of peak is quite sensitive to the cluster  (refered as A in \cite{fi}) mass.
But the probability for the mass of the cluster around the proper value is very small.
So explaining the large `production rate'
needs fine tuning by  the final state interaction models.

3) \texttt{Combination model}.
 The key thing is how the light flavour produced and combined with the bottom quark,
 to become X(5568).
 The  probability of creation of each flavour is as the above, only we do not need to introduce the
 diquark. So it is the $b$ quark combined with $\bar s u \bar d$  or $\bar s d \bar u$ for X(5568).
 If we assume that the total probability of any three light  (anti)quarks combined with $ b$  as 1, then
 \begin{equation}
 P_X=\frac{\lambda}{D} \frac{1}{D} \frac{1}{D} \times 2.
 \end{equation}
Here $D=2(1+1+\lambda)$.
If we assume that $b$ combines with any other antiquark with probability 1, then
\begin{equation}
 P_{Bs}=\frac{\lambda}{(D/2)},
  \end{equation}
\begin{equation}
 \rho^{max}_{combination}= \frac{P_X}{P_{Bs}} \sim 5 \%.
 \end{equation}
  Here we write the superscript `$^{max}$'  to address the fact:
   We did not take into account that the above two cases,
  $b$ combining with an antiquark or $b$ combining with three quark cluster $q \bar q' \bar q''$ excludes
  each other, and there should be other cases, e.g., $b$ combining  with  $qq$ to form a baryon. All these possibilities
  add up to 1.
   This reflects the unitarity of the combination process \cite{long1, lsy}, i.e., one quark
  can combine with many other quarks or clusters, according to some law, but the total probability must be 1, since the
   quark is confined and has to go into a hadron. For the calculation of $\rho$ here, we can introduce a relative probability
  $\zeta = (b ~ combines ~ with~any ~q \bar q' \bar q'')/ (b~ combines ~with ~ any ~ \bar q)$. In general $\zeta$ is smaller than 1, according to the fact that four quark state is not copiously found
  in hadronization. Hence $\rho_{combination}=\rho^{max}_{combination} \times \zeta$, smaller than $5\%$.





 The above analysis is dependent on the SU(3) flavour symmetry breaking parameter $\lambda$, with different sensitivities
 for string model and combination model, respectively.  In concrete, the result of the combination model is less sensitive to $\lambda$, our up
 limit varies from about $3 \%$ to $6 \%$ for $\lambda $ from 1 to 0. The result of the string model is much more sensitive,
 varies from 0 to about  $16 \%$ for $\lambda $ from 0 to 1. Especially, if we want to get $8\%$ in the string model,
 we need to take $\lambda $ to be 0.65.  
 However, $\lambda$ is a very steady value, and corresponds to  the physics of producing a strange quark pair from the vacuum
 by tunneling effect in the hadronization process, independent from  colliding particles (except heavy nuclei) and energies. It is taken as 0.3 in Pythia \cite{Pythia}, and  confirmed by experiments up to LHC energies \cite{Abelev:2012tca}.
 
 Though the above results can be considered consistent with data at the order of
magnitude level, the tension is obvious. 
 Since no other plausible parameters to tune, one has difficulty to raise to a larger production ratio (if confirmed by experiment) in these models.     
  %
  %
 %
   On the other hand, 
the combination model   has provided a clue
to improve the description on the large production ratio. 
 If there is a special large $\zeta_{X}$ which is only applicable to
this special X(5568) production, it can provide an enhancement
factor. In the combination  model, if  it can take the value around 2, then we can get the experimental result.
In  the mean time, for the measurement \cite{D0:2016mwd},   $B_s K$ and $B_d \pi$ are searched as cross check and no signal of new state found \cite{epaps}.
It seems another evidence implying that the X(5568) is a very special structure. Its production
may be quite unique, different from the other particles produced, though QCD is flavour blind. So this  plays another
 support for us
 to deal it independently, ignoring the constrains from the  unitarity as well as the general hadronization
 mechanism applied to the general hadrons.
 For other cases, the $\zeta$ is quite small,
  as the ansatz for the string model, a value
  of order of magnitude 10 per cent, so even the other combination can  lead to a 4-quark state but no
   observable signal \cite{long1}.

%



The above idea of special $\zeta_X$ can be more systematically realized in analogy of
 the inclusive resonance production framework, as that for the quarkonium. 
The key point is to describe its production in two steps, whatsoever
taking it as a bound state of hadrons or 2-quark clusters,
or even, two pieces of strings.  
  And since this special production mechanism, the probability is not
 added with others to exhaust the unitarity constrain \cite{long1, lsy, Wang:2016vxa} mentioned above, besides not applied to other
 light flavours.





As \cite{long1}, we start from the amplitude
\begin{equation}
A(P)  = <H(A,B), X|\hat{T}|p \bar p >  =  \frac{1}{\sqrt{\mu}} \int \frac{d^3 k}{(2\pi)^3} \Phi(\vec k)
           {\bold M}(\vec k).  
 \end{equation}
$\mu$ is the reduced mass. A and B are two clusters to be combined as the X(5568). Here we take the
$p \bar p$ collision process as the example.

The relative momentum $q$ of these two clusters can be considered as small in the rest frame of X(5568),
we get
\begin{equation}
\label{swave}
A (P)= \frac{1}{\sqrt{\mu}} \Psi(0)  \hat{O}(q=0), q=P_A-P_B.
\end{equation}
Here  $\Psi$ is the wave function.  $\hat{O}$ is  the  amplitude of production of two free ingredient particles
(with vanishing relative momentum and proper angular momentum state).
We only consider the simplest S-wave case.
For the cross section,  we need the absolute value square, with the proper initial flux factor $1/F$ and phase space integral.

From this formula,  the X(5568) is produced in two steps:  First,  the production of the ingredient
hadron/cluster pair; second, the combination of this pair to X(5568) with probability described by
$|\Psi(0)|^2$.
For the first step, it is the
\begin{eqnarray}
	\label{int}
&& \frac{1}{F} \sum_{j \neq A, B} \hspace{-0.67 cm} \int \prod \frac{d^3 p_j}{(2 \pi)^3 2 E_j} \overline{ |\hat{O} |^2} (p_j,  P_A+P_B=P_H, q=0) \nonumber \\
&& \times (2 \pi)^4 \delta^{(4)}(P_{intial}-\sum_{j \neq A, B} p_j-P_H)
\end{eqnarray}
to be calculated. Here the average is  taken on various spin states.
It is not possible to be calculated directly with some effective quantum field theory/model when the initial
state is (anti) protons and A and B are hadrons or diquarks.  However,
\begin{eqnarray}
	\label{int2}
&& \frac{1}{N} \frac{d N}{d^3 P_H d^3 q} \\
&& \propto \frac{1}{F} \sum_{j \neq A, B} \hspace{-0.67 cm} \int \prod \frac{d^3 p_j}{(2 \pi)^3 2 E_j} \overline{ |\hat{O} |^2} (p_j,  P_A+P_B=P_H, q) \nonumber \\
&& \times (2 \pi)^4 \delta^{(4)}(P_{intial}-\sum_{j \neq A, B} p_j-P_A-P_B) \nonumber
\end{eqnarray}
  can be calculated by an event generator such as Pythia \cite{Pythia}
 or equivalently SDQCM \cite{Jin:2010wg} for the case that hadrons/diaurks A and B  on shell.
 It is the advantage that in the framework we employ, only the on shell case is considered, so that
 the numerical calculation with event generator is plausible.
 %
  The quantity of Eq. (\ref{int2}) describes the two hadrons/diquarks A and B correlation in the phase space. For the hadron case, by proper integral on components of  $P_H$ and/or q, the resulting correlations can directly be compared with
 data and    serve for tuning the parameters.


 Employing the event generator, one gets
 \begin{equation} \label{ext}
   \frac{1}{N} \frac{d N}{d^3 P_H d^3 q}, \forall q,
 \end{equation}
then extrapolates to the special case $q=0$. Numerically, one can take an average around $q=0$ for the above
quantity \cite{long1}.


For  $B_s \pi$, we
take the simple average of two cases,  hadron bound state (B, K) and diquark pair as mentoned above,
then  fit the X(5568) spectrum measured by D0 Collaboration to get the effective wave function at origin (Fig. 1(a)).
\begin{figure}
\includegraphics[scale=0.02]{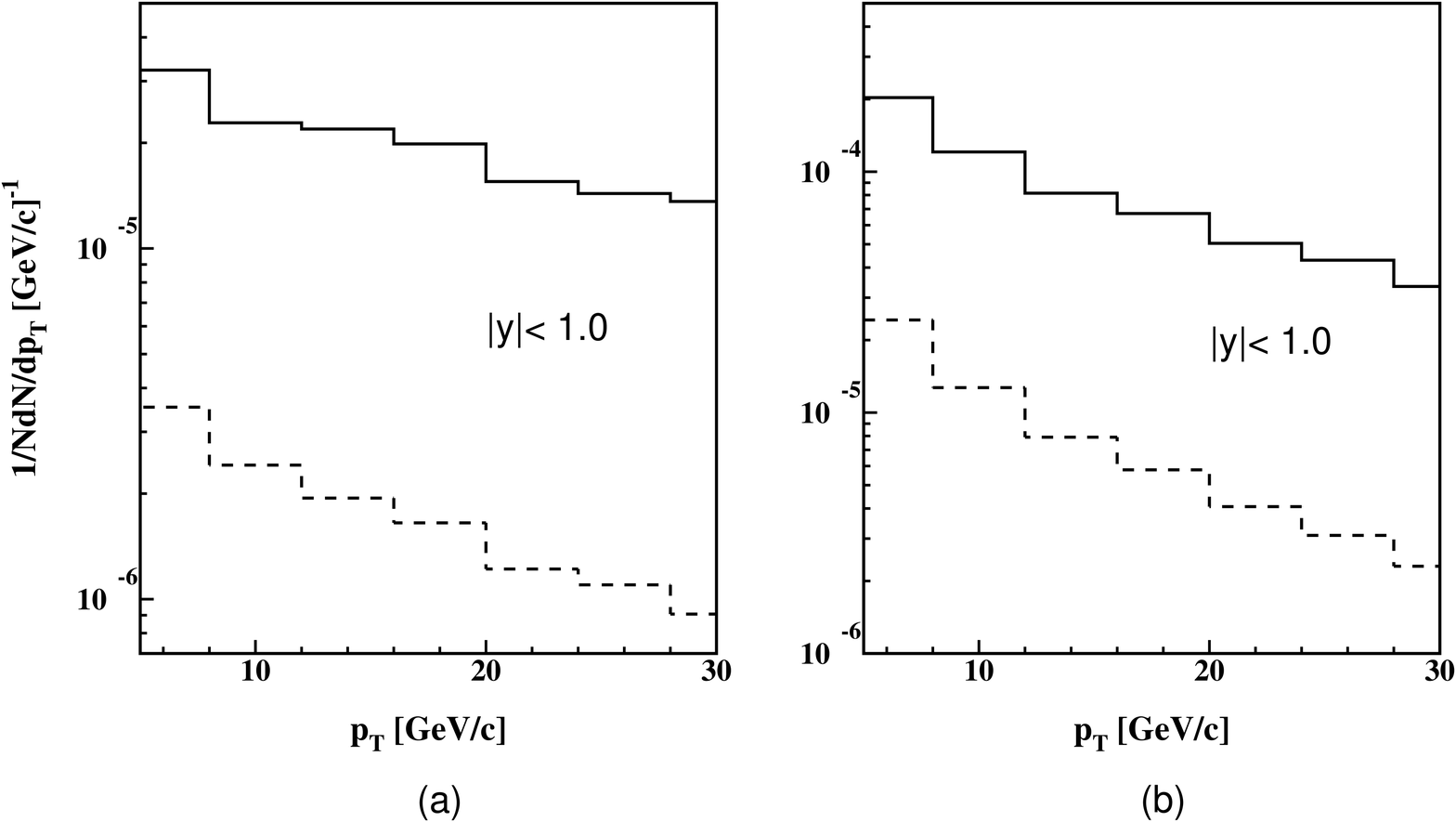}
\caption{Transverse momentum distributions at Tevatron. (a) The dashed line is that for X(5568), with the best fitting of the wave function  to get the
correct $\rho$ measured by D0 collaboration. The solid line is for $B_s$ as reference.  (b) For the charm sector, dashed  - X$_c(D_s ^\pm \pi^\pm)$;
solid -  $D_s$. }
\label{bs}
\end{figure}
%
%
 The X(5568) transverse spectrum is  softer than that of  $B_s$,  as demonstrated here
and indicated from the experiment.
 This is from the fact that we require
 the two clusters near to each other in phase space for  combination. Realized in the above formulation, is the  relative momentum vanishing. This is in contrary to the fragmentation spectrum, the more massive, the harder.


The cross sections of X(5568) in other collision processes and energies are easy to be obtained, since the effective
wave function at origin  is process- and energy- independent.  Here we show the
 pseudo-rapidity $\eta$ distributions  for proton-proton collisions at $\sqrt{s}=8$ TeV
as an example (Fig. 2(a)). For others we refer to \cite{long1}.  The production rate begins to fall beyond $\eta =3$, as general B hadrons. The  `rapidity plateau' is much more narrow than those of  light charged hadrons (mainly pions).

These results are useful  for various detectors.
Based on our calculation, one can  go further to estimate the kinematic distributions of the  signal particles  which are
from the  decay of X(5568) and  directly detected.
As an example, Fig. 2 (b) showes the $k_T-k$ (transverse momentum and  total momentum) distribution for the
signal pions from the decay process $X \to B_s + \pi$ in the LHCb detector ranges ($2< \eta <5$).
The mass difference between X(5568) and $B_s+\pi$ is  small,  and the  pion mass is small. These facts lead to that
the produced pions are not energetic,  e.g.,  only around  $10 \%$ of the signal pions with $k_T > 0.5$ GeV/c
 (the requirement of the relevant measurement by LHCb Collaboration \cite{LHCb:2016ppf}).

The formulation is also applicable for the cross section of
X$_c(D^{\pm}_s\pi^{\pm}$) state production.
Both charm and bottom are  heavy, and can be calculated by perturbative QCD
in the exactly same way once taking into account the different value of the mass.
If  $X_c$ exists, it is not
difficult to be detected.
$D^{\pm}_s$ can be detected from $D^{\pm}_s \to \phi \pi$ channel, by proper 3 charged particle tracks from the vertex
displaced from the primary one.
Then this  reconstructed  $D^{\pm}_s$ can be combined with a proper  charged particle track considered as $\pi$ from the {\it primary vertex}
to give  the invariant mass distribution to look for the resonance.
If $K^0_s$ is well measured, $D^{\pm}_s$ can also be reconstructed from the 2K channel and then combined with the
$\pi$ from the primary vertex.
This kind of pions can eliminate the possibility that the $X_c$ produced from the decay of bottom.
Of course just by keeping or not this restriction, one can preliminarily investigate $X_c$ from multi-production
or from weak decay.
Here we would like to emphasize that, since the mass of $X_c$ are around half of X(5568), it has a larger boost factor $\gamma$
about two times of that of X(5568) for the same momentum. This means whether Tevatron or LHC,  in both central and large
rapidity regions,  the signal pions are more energetic to be detectable.


The transverse distribution of $X_c$ at Tevatron energy, comparing with that of $D_s$,  can be seen from Fig. \ref{bs} (b). Here we assume that  replacing $b$ by $c$ quark will not change the value of the wave function at origin, since the reduced mass is insensitive to the heavy ingredient. The  production ratio
$\rho_c=X_c/D_s$  for the transverse momentum region 10 $< p_T (D_s) <$ 15 GeV/$c$ and 15 $< p_T (D_s) <$ 30 GeV/$c$, is $ 10.2 \%$ and $ 7.9 \%$, respectively, almost similar as those of the bottom   
sector. It is  copious enough and the search from experiments is reasonable.




\begin{figure}[htb]
	\centering
	\begin{tabular}{cc}
		\scalebox{0.10}{\includegraphics{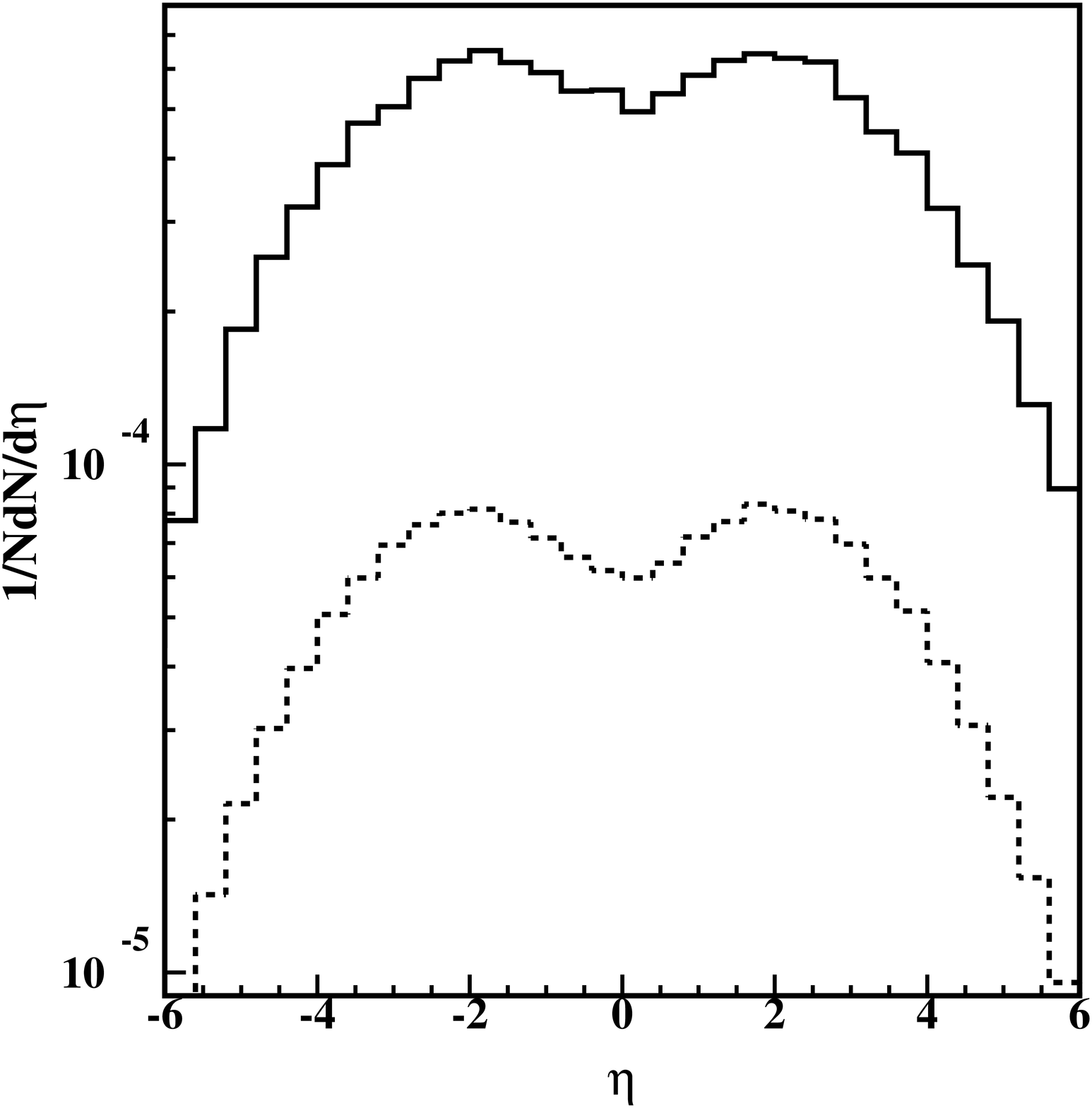}}&
		\scalebox{0.10}{\includegraphics{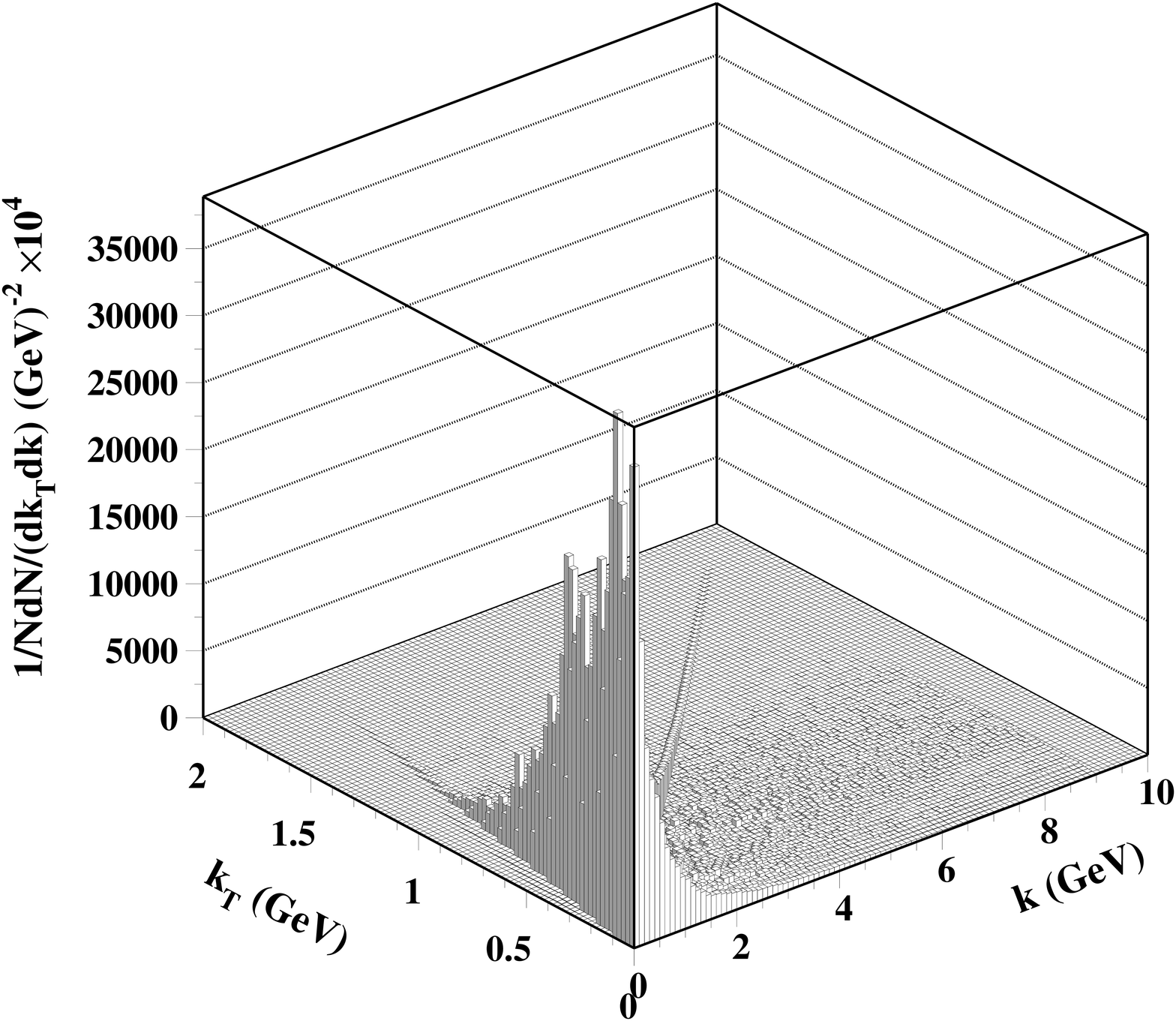}}\\
		{\scriptsize (a)}&{\scriptsize (b)}
	\end{tabular}
	\caption{
		(a) Pseudo-rapidity distributions. The dashed line is for X(5568), the solid is for $B_s$.
        (b)  $k_T-k$ (transverse momentum and  total momentum) distribution of the signal pions. A Breit-Wigner
        form of the X(5568) mass distribution is convoluted  ($\Gamma_X=21.9$ MeV/$c^2$) \cite{D0:2016mwd}.}
	\label{lhcby}
\end{figure}

We look forward for  the new charm four quark partner. This  will shed light to the understanding of  this special
new state,
 as well as deepen our understanding on the hadronization mechanism, besides its structure \cite{rw}. 
These productions of X(5568) or possible $X_c$ can also be realized in high energy heavy ion collisions, with a larger rate since there strangeness and/or diquark is enhanced.
The large number of quarks in unit phase space volume in heavy ion collisions also indicates that one can get a larger value of
$\hat{O}$.  These lead to larger production ratio $X(5568)/B^0_s$ ($X_c/D_s$), hence larger $B^0_s/B^0$ from X(5568) decay
(also $D_s/D$). This can be measured as the `anomalous strangeness  enhancement' in the B (D) meson sector.
Furthermore, it is also interesting to combine X(5568)   with other tracks to look for heavier hadron, e.g.,  $B_c$,
as complementary study.
If one finds  an $X_c$, to study  those  from multi-production  and from heavier hadron decay are also very
helpful, as mentioned above.





We thank Profs.  T Gershon,   Y. R. Liu,  L.L. Ma and Z. G. Si  for discussions. This work is supported in part by NSFC and NSF, Shandong Province.

\end{document}